\documentclass[english,submission]{article}
\usepackage{graphicx}
\usepackage[utf8]{inputenc}
\usepackage[english]{babel}
\usepackage{url}

\begin{document}

%\paperdetails{
%  perspective=sciencetheoretical,
%  area={Distributed systems programming}
%}

\title{On the Design of Distributed Programming Models}
\author{Christopher S. Meiklejohn \\ christopher.meiklejohn@gmail.com}
%\affiliation{Universit\'e catholique de Louvain, Louvain-la-Neuve, Belgium}

% \keywords{distributed systems, consensus, eventual consistency}

%\begin{CCSXML}
%<ccs2012>
%<concept>
%<concept_id>10002944.10011122.10003459</concept_id>
%<concept_desc>General and reference~Computing standards, RFCs and guidelines</concept_desc>
%<concept_significance>300</concept_significance>
%</concept>
%</ccs2012>
%\end{CCSXML}

%\ccsdesc[300]{General and reference~Computing standards, RFCs and guidelines}

\maketitle

\begin{abstract}
Programming large-scale distributed applications requires new abstractions and models to be done well.  We demonstrate that these models are possible.

Following from both the FLP result and CAP theorem, we show that concurrent programming models are necessary, but not sufficient, in the construction of large-scale distributed systems because of the problem of failure and network partitions: languages need to be able to capture and encode the tradeoffs between consistency and availability.

We present two programming models, Lasp and Austere, each of which makes a strong tradeoff with respects to the CAP theorem.  These two models outline the bounds of distributed model design: strictly AP or strictly CP.  We argue that all possible distributed programming models must come from this design space, and present one practical design that allows declarative specification of consistency tradeoffs, called Spry.
\end{abstract}

\section{Introduction}

Languages for building large-scale distributed applications experienced a Golden Age in the 1980s with innovations in networking and the invention of the Internet.  These language tried to ease the development of these applications, influenced by the growing requirements of increased computing resources, larger storage capacity, and the desired high-availability and fault-tolerance of applications.

Two of the most widely known from the era, Argus~\cite{liskov1988distributed} and Emerald~\cite{black2007development} each took a different approach to solving the problem, and the abstractions that each of these languages provided aimed to simplify the creation of correct applications, reduce uncertainty when dealing with unreliable networks, and alleviate the burden of dealing with low-level details related to a dynamic network topology; Emerald focusing heavily on object mobility and Argus on atomic transactions across objects residing on multiple machines. 

As it stands, these languages never saw any adoption\footnote{One notable exception here is the Distributed Erlang~\cite{wikstrom1994distributed} extension for the Erlang programming model, later adopted by Haskell~\cite{epstein2011towards}.} and most of the large-scale distributed applications that exist today have been built with sequential or concurrent programming languages such as Go, Rust, C/C++, and Java.  These languages have taken a library approach to distribution, adopting many ideas from languages such as Emerald and Argus.  We can highlight two examples: first, the concept of \textit{promises} from Argus, is now a standard mechanism relied upon when issuing an asynchronous request~\cite{eriksen2013your, liskov1988promises, syme2011f}; second, from Emerald, the concept of a directory service that maintains the current location of mobile processes~\cite{bykov2011orleans}.

Distributed programming today has become ``the new normal.''  Nowadays, whether you are building a mobile or a rich web application, developers are increasingly trying to provide a ``near native'' experience~\cite{charland2011mobile}, where application users feel as if the application is running on their machine.  To achieve this, shared state is usually replicated to devices, locally mutated and periodically synchronized with a server.  In these scenarios, consistency can become increasingly challenging if the application is to stay available when the server cannot be reached.  \textbf{Therefore, it is now paramount that we have tools for building correct distributed applications.}

We argue that the reason that these previous attempts at building languages for large-scale distributed systems have failed to see adoption is that they fail to capture the requirements of today's application developers.  For instance, if an application must operate offline, a language should have primitives for managing consistency and conflict resolution; similarly, if a language is to adhere to a strict latency bound for each distributed operation, the language should have primitives for expressing these bounds.  

In this paper, we relate these real-world application requirements to the CAP theorem~\cite{brewer2000towards, gilbert2002brewer}, showing that a distributed application must sacrifice consistency if it wishes to remain available under network partitions.  We demonstrate that there are several design points for programming models that could, and do, exist within the bounds of the CAP theorem, with two examples that declaratively specify application-level distribution requirements.

\section{Sequential and Concurrent Programming}

We explore the challenges when moving from sequential programming to concurrent programming.

\subsection{Sequential Programming}

Most of the programming models that have widespread adoption today are designed in von Neumann style\footnote{Functional and logic programming remain notable exceptions here, although their influence is minimal in comparison.}: computation revolves around mutable storage locations whose values vary with time, aptly called variables, and control statements are used to order assignment statements that mutate these storage locations.  Programs are seen to progress in a particular order, with a single thread of execution, and terminate when they reach the end of the ordered list of statements.

\subsection{Concurrent Programming}

Concurrent programming extends sequential programming with multiple sequential threads of execution: this is done to leverage all available performance of the computer by allowing multiple tasks to execute at the same time.  Each of these threads of execution could be executing in parallel, if multiple processors happened to be available, or a single processor could be alternating control between each of the threads of execution, giving a certain amount of execution time to each of the threads.  

Concurrent programming is difficult.  In the von Neumann model where shared memory locations are mutated by the sequential threads of execution, care must be taken to prevent uncontrolled access to these memory locations, as this may sacrifice correctness of the concurrent program.  For instance, consider the case where two sequential threads of execution, executing in parallel, happen to read the same location and write back the location incremented by $1$.  It is trivial to observe that without controlled access to the shared memory location, both threads could increment the location to $2$; effectively ``losing'' one of the valid updates to the counter.

Originally described by Dijkstra~\cite{dijkstra2001solution}, this ``mutual exclusion'' problem is the fundamental problem of concurrent computation with shared memory.  How can we provide a mechanism that allows correct programming where multiple sequential threads of execution can mutate memory that is visible by all nodes of the system.  Dijkstra innovated many techniques in this area, but the most famous technique he introduced was that of the ``mutex'' or ``mutual exclusion.''

Mutual exclusion is the process of acquiring an exclusive lock to a shared memory location to ensure safe mutation.  Returning to our previous example, if each sequential thread of execution was to acquire an exclusive lock to the memory location before reading and updating the value, we no longer have to worry about the correctness of the application.  However, mutual exclusion can be difficult to get right when multiple locks are required, if they are not handled correctly to avoid deadlock.

%% If each of our threads of execution was scheduled on a single processor, and we happened to de-schedule the first thread of execution after it has acquired the lock, but before we had released the lock, the second thread of execution would not be able to proceed and our application would deadlock.  Therefore, programming abstractions aim to ease the development of concurrent applications and reduce the programmer burden on making sure these locks are handled in a principled manner.

But, how is the programmer to reason about whether their concurrent application has been programmed correctly?  Given multiple threads of execution, and a nondeterministic scheduler, there exists an exponential number of possible schedules, or executions, the program may take, where programmers desire that each of these executions result in the same value, regardless of schedule.

Therefore, most programmers desire a correspondence commonly referred to as confluence.  Confluence states simply, that the evaluation order of a program, does not impact the outcome of the program.  In terms of the correspondence, programmers ideally can write code in a sequential manner, that can be executed concurrently, and possibly in parallel, and any of the possible schedules that it may take when executed, results in the same outcome of the sequential execution.

From a formal perspective, we have several correctness criteria for expressing whether a concurrent execution was correct.  For instance, sequential consistency~\cite{lamport1979make, goodman1991cache} is a correctness criteria for a concurrent program that states that the execution and mutation of shared memory reflects the program order in the textual specification of the program; linearizability~\cite{herlihy1990linearizability} states that a concurrent execution followed the real time order of shared memory accesses and strengthens the guarantees of sequential consistency. 

\section{Distributed Programming}

Distributed programming, while superficially believed to be an extension of concurrent programming, has its own fundamental challenges that must be overcome.

\subsection{The Reasons For Distribution}

Distributed programming extends concurrent programming even further.  Given a program that's already concurrent in it's execution, programmers distribute the sequential threads of execution across multiple machines that are communicating on a network.  Programmers do this for several reasons: 

\begin{itemize}

\item \textbf{Working set.} The working data set or problem programmers are trying to solve will take too long to execute on a single machine or fit on a single machine in memory, and therefore programmers need to distribute the work across multiple machines.

\item \textbf{Data replication.} Programmers need data replication, to ensure that a failure of one machine does not cause the failure of our entire application.

\item \textbf{Inherent distribution.} Programmers application's are inherently distributed; for example, a client application living on a mobile device being serviced by a server application living at a data center.

\end{itemize}

What programmers have learned from concurrent programming is that accesses to shared memory should be controlled: programmers may be tempted to use the techniques of concurrent programming, such as mutexes, monitors~\cite{hoare1974monitors}, and semaphores, to control access to shared memory and perform principled, safe, mutation.

\subsection{The Challenges of Distribution}

On the surface, it appears that distribution is just an extension of concurrent programming: we have taken applications that relied on multiple threads of execution to work in concert to achieve a goal and only relocated the threads of execution to gain more performance and more computational resources.  However, this point of view is fatally flawed.

As previously mentioned, the challenges of concurrent programming are the challenges of nondeterminism.  The techniques pioneered by both Dijkstra and Hoare were mainly developed to ensure that nondeterminism in scheduling did not result in nondeterminism in program output.  Normally, we do not want the same application, with fixed inputs, to return different output values across multiple executions because the scheduler happened to schedule the threads in different orders.

Distribution is fundamentally different from concurrent programming: machines that communicate on a network may be, at times, unreachable, completely failed, unable to answer a request, or, in the worst case, permanently destroyed.  Therefore, it should follow that our existing tools are insufficient to solve the problems of distributed programming.  We refer to these classes of failures in distributed programming as ``partial failure'': in an effort to make machines appear as a single, logical unit of computation, individual machines that make up the whole may independently fail.

Distributed systems researchers have realized this, and have identified the core problem of distributed system as the following: the agreement problem.  The agreement problem takes two forms, duals of each other, mainly:

\begin{itemize}

\item \textbf{Leader election.} The process of selecting an active leader amongst a group of nodes to act as a sequencer or coordinator of operations for those nodes; and

\item \textbf{Failure detection.} The process of detecting a node that has failed and can no longer act as a leader.

\end{itemize}

These problems, and the problems of locking, are only exacerbated by two fundamental impossibility results in distributed computing: the CAP theorem~\cite{gilbert2002brewer} and the FLP result~\cite{fischer1985impossibility}.

The FLP result demonstrates that on a truly asynchronous system, agreement, when one process in the agreement process has failed, is impossible.  To make this a bit more clear, when we can not determine how long a process will take to perform a step of computation, and we can not determine how long a message will take to arrive from a remote party on the network, there is no way to tell if the process is just delayed in responding or failed: we may have to wait an arbitrarily long amount of time.  

%We place timeouts on how long a process should take to respond, and if the timeout is exceeded, we just assume that it's failed.  This can be problematic if this node returns later and attempts to contribute to the computation process. 

%% TODO from AJS:
%% Also, in the previous paragraph there is really no mention of the problem that this solution fixes. The randomized timeouts really only fix the problem of each node trying to start a new election concurrently with other nodes and cancelling each others attempts. FLP still allows this randomization to fail to correct infinite elections if message delays are also random and tend to synchronize with the timeouts by chance. However in practice, delays aren't random and the probability of the delays synchronizing with the timeouts is very small.

FLP is solved in practice via randomized timeouts that introduce nondeterminism into the leader election process to prevent infinite elections.  Algorithms that solve the agreement protocol, like the Raft consensus protocol~\cite{ongaro2014search} and the Paxos leader election algorithm~\cite{lamport1998part, chandra2007paxos} take these timeouts into account and take measures to prevent a seemingly faulty leader from sacrificing the correctness of a distributed application.

The CAP theorem states another fundamental result in distributed programming.  CAP states simply, that if we wish to maintain linearizability when working with replicated data, we must sacrifice the ability for our applications to service requests to that shared memory if we wish to remain operational when some of the processes in the system can not communicate with each other.  Therefore, for distributed applications to be able to continue to operate when not all of the processes in the system are able to communicate -- such as when developing large-scale mobile applications or even simple applications where state is cached in a web browser -- we have to sacrifice safe access to shared memory.

Both CAP and FLP incentivize developers to avoid using replicated, shared state, if that state needs to be synchronized to ensure consistent access to it.  Applications that rely on shared state are bound to have reduced availability, because they either need to wait for timeouts related to failure detection or for the cost of coordinating changes across multiple replicas of the shared state.

\section{Two Extremes in the Design Space}

While development of large-scale distributed applications with both sequential and concurrent programming models has been widely successful in industry, most of these successes have been supported by systems that close the gap between a language that has distribution as a first-class citizen, and a concurrent language where tools that solve both failure detection and the agreement problem are used to augment the language.

For programming models to be successful for large-scale distributed programming, they need to embrace the tradeoffs of both the FLP result and the CAP theorem.  We believe that there exists a space, in what we refer to as the boundaries of the CAP theorem, where a set of programming models that take into account the tradeoffs of the CAP theorem, can exist and flourish as systems for building distributed applications.

We now demonstrate two extremes in the design space.  First, Lasp, a programming model that sacrifices consistency for availability.  Second, Austere, a programming model that sacrifices availability for consistency.  Both of these models  sit at extreme sides of the spectrum proposed by the CAP theorem.

Both of these languages share a common design component: a datastore tracking local replicas of shared state, or ``variable'' state.  To ensure recency of these replicas, a propagation mechanism is used: where strong consistency is required, this protocol may be driven by a consensus protocol such as Raft~\cite{ongaro2014search} or Paxos~\cite{lamport1998part}; where weaker consistency is required, a simple anti-entropy protocol~\cite{demers1987epidemic} may suffice.  Where the models differ is in there evaluation semantics.  Application written in these models may choose to synchronize replicas before using a value in a computation or not, depending on whether the model prefers availability or consistency.

Each of these models is a small extension to the $\lambda$-calculus.  This extension provides named registers pointing to location in the data store.  These locations designate their primary location and data type, and are dereferenced during substitution.  In the event the replica has to be refreshed before evaluation, delimited continuations~\cite{felleisen1988theory, kiselyov2007delimited} are used as a method of interposition to insert the required synchronization code, managed by a scheduling thread.  This same mechanism is used to periodically refresh values in the background either through anti-entropy sessions or consensus.

\subsection{Lasp}

Lasp~\cite{meiklejohn2015lasp, meiklejohn2015selective} is a programming model designed as part of the SyncFree and LightKone EU projects~\cite{syncfree, lightkone} focusing on synchronization-free programming of large-scale distributed applications.  Lasp sits at one extreme of the CAP theorem: Lasp will sacrifice consistency in order to remain available.

%% CRDT

Lasp's only data abstraction is the Conflict-Free Replicated Data Type (CRDT)~\cite{shapiro2011conflict}.  A CRDT is a replicated abstract data type that has a well defined merge operation for deterministically combining the state of all replicas, and Lasp builds upon one specific variant of CRDTs: state-based CRDTs.  CRDTs guarantee that once all messages are delivered to all replicas, all replicas will converge to the same result. 

With state-based CRDTs\footnote{Herein referred to as just CRDTs.}, each data structure forms a bounded join semilattice, where the join operation computes the least-upper-bound for any two elements.  While CRDTs come in a variety of flavors, like sets, counters, and flags (booleans), two main things must be keep in mind when specifying new CRDTs:

\begin{itemize}
\item CRDTs are replicated, and by that fact inherently concurrent.  Therefore, when building a CRDT version of a set, the developer must define semantics for all possible pairs of concurrent operations: for instance, a concurrent addition and removal of the same element.
\item To ensure replica convergence with minimal coordination, it follows from the join-semilattice that the join operation compute a least-upper-bound: therefore, all operations on CRDTs must be associative, commutative, and idempotent.
\end{itemize}

%% Lasp

Lasp is a programming model that allows developers to do basic functional programming with CRDTs, without requiring application developers to work directly with the bounded join-semilattice structures themselves: in Lasp, a developer sees a CRDT set as a sequential set.  Given all of the data structures in Lasp are CRDTs themselves, the output of Lasp applications are also CRDTs that can be joined to combine their results.

Lasp never sacrifices availability: updates are always performed against a local replica and state is eventually merged with other replicas.  Consistency is sacrificed: while replica convergence is ensured, it may take an arbitrarily long amount of time for convergence to be reached and updates may arrive in any order.

However, Lasp has several strong restrictions given the CRDT foundation that provides its availability: all operations must commute and all data structures must be able to be expressed as a bounded join-semilattice.  This obviously rules out several common data structures, such one very important one: the list.

\subsection{Austere}

Austere is a programming model where all replicated, shared state is synchronized for every operation in the system.  Austere sits at another extreme of the CAP theorem: Austere will sacrifice availability in order to preserve consistency.

Before any access or modification to replicated, shared state, Austere will contact all replicas using two-phase locking (2PL)~\cite{lausen2009two} to ensure values are read without interleaving writes with communication, and two-phase commit (2PC)~\cite{al2009two} to commit all modifications.  In the event that a replica can not be contacted, the system will fail to make progress, ensuring a single system image: reducing distributed programming to a single sequential execution to ensure a consistent view across all replicas.

Compared to Lasp, Austere shares the common $\lambda$-calculus core; however, the gain of semantics in Austere is paid for by  reduced availability.

\section{``Next 700'' Distributed Programming Models}

\begin{figure}[h]
\begin{center}
\noindent\includegraphics[scale=0.75]{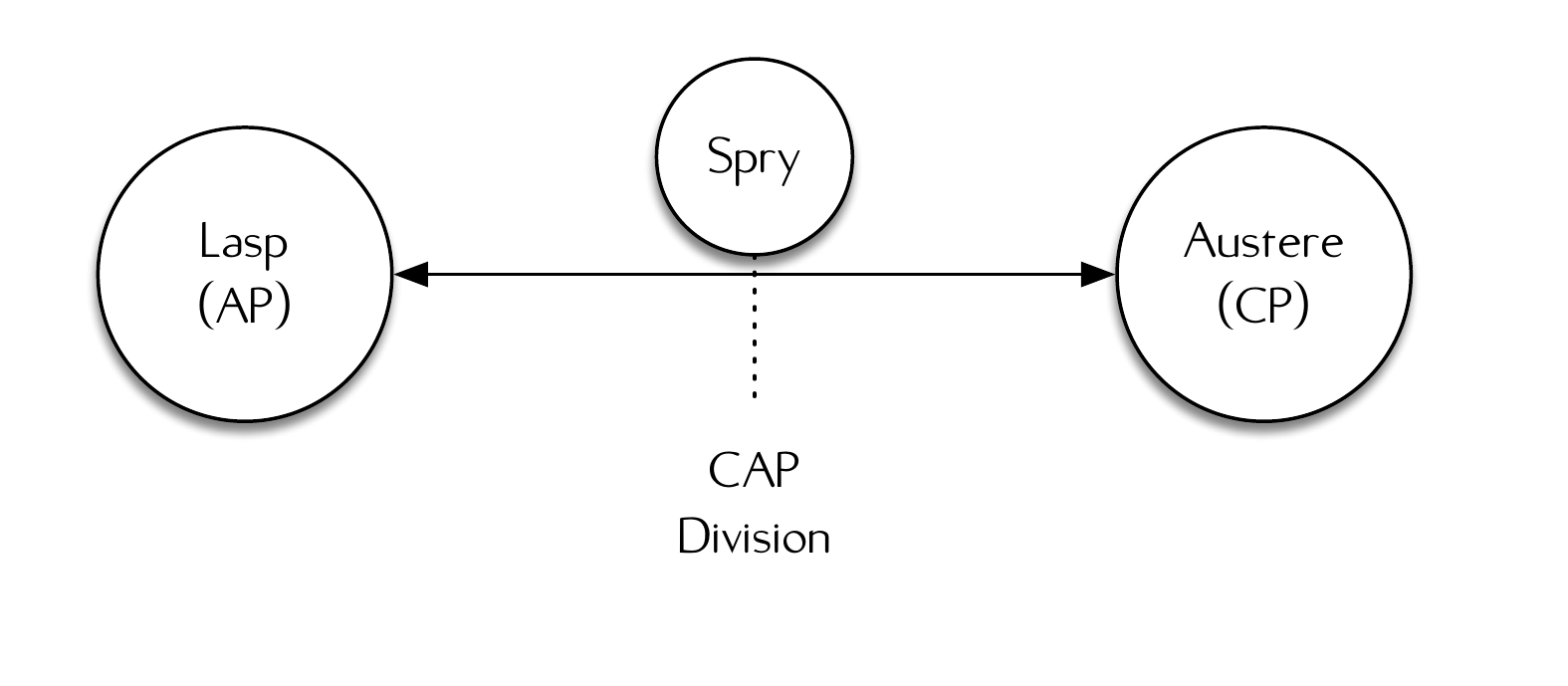}
\end{center}
\caption{The design space of the ``next 700'' distributed programming models.}
\label{fig:models}
\end{figure}

The ``next 700''~\cite{landin1966next} distributed programming models will set between the bounds presented by Austere and Lasp: extreme availability where consistency is sacrificed vs. extreme consistency where availability is sacrificed. (see Figure~\ref{fig:models})

More practically, we believe that the most useful languages in this space will allow the developer to specifically trade-off between availability and consistency at the application-level.  This is because the unreliability of the network, dynamicity of real networks, and production workloads require applications to remain flexible.  These trade-offs should be specified declaratively, close to the application code.  Application developers should not have to reason about transaction isolation levels or consistency models when writing application code.

We provide an example of one such language, Spry, that makes a trade-off between availability and consistency based on application-level semantics.  We target this language for one use case in particular: Content Delivery Networks (CDNs), where application-level semantics can be used to declaratively specify Service Level Agreements (SLAs.)

\subsection{Spry}

Spry~\cite{spry} is a programming model for building applications that want to tradeoff availability and consistency at varying points in application code to support application requirements.

We demonstrate two different types of tradeoffs that application developers might make in the same application.  Consider the case of a Content Delivery Network (CDN), an extremely large-scale distributed application.

\begin{itemize}
\item \textbf{Availability for consistency.}  In a CDN, the system tries to ensure that content that is older than a particular number of seconds will never be returned to the user.  This is usually specified by the application developer explicitly, by checking the object's staleness and fetching the data from the origin before returning the response to the user.
\item \textbf{Consistency for availability.}  CDN's usually maintain partitioned inverted indexes that can be queried to search for a piece of content.  Because nodes may become unavailable, or respond slowly because of high load, application developers may want to specify that a query returns results from a local cache if fetching the index over the network takes more than a particular amount of time.  This is usually specified by the application developer explicitly, by setting a timer, attempting to retrieve the object over the network, reusing cached results if the latency bound can not be met, and returning the response to the user.
\end{itemize}

Application developers specify these constraints declaratively in Spry.  If a replicated value should not be older than a particular number of milliseconds, developers can annotate these values with the bounded staleness requirements.  If a replicated value should always be as fresh as it can be within a bound of a number of milliseconds, this can be specified as well.  Similarly, these values can be tweaked while the application is running, allowing developers to adjust the system while it is operating, responding to failures or increased load.

\section{Conclusion}

We have seen that the move from sequential programming to concurrent programming was fairly straightforward: all that was required was a principled approach to state mutation through the use of techniques like locking to prevent values from being corrupted, which can lead to unsafe programs.  However, the move to distributed programming is much more difficult because of the uncertainty that is inherent in distributed programming.  For example: will this machine respond in time?  Has this machine failed and is it able to respond?  

Distributed programming is a different beast, and we need programming models for adapting accordingly.  Can we come up with new abstractions and programming models that aid in expressing the application developers intent \textbf{declaratively?}

We believe that it is possible.  We have shown you three different programming model designs that all make different tradeoffs when nodes become unavailable.  Two of these, Lasp and Austere, provide the boundaries in model design that are aligned with the constraints of the CAP theorem.  One of these, Spry, takes a declarative approach that puts the tradeoffs in the hands of the application developer.  All three of these can cohabit the same underlying concurrent language.

\section*{Acknowledgements} 
We want to thank Zeeshan Lakhani, Justin Sheehy, Andrew J. Stone, and Zach Tellman for their feedback.

\bibliography{cap-space-article.bib}
\bibliographystyle{abbrv}
\end{document}